\documentclass[11pt,oneside,a4paper]{book}

\usepackage[T1]{fontenc}
\usepackage[utf8]{inputenc}
\usepackage{lmodern}
\usepackage{microtype}
\usepackage[a4paper,margin=1.2in]{geometry}
\usepackage{amsmath,amssymb,mathtools}
\usepackage[hidelinks]{hyperref}
\usepackage{xstring}
\usepackage{graphicx}
\usepackage{float} 

\DeclareMathOperator{\MI}{I}
\DeclareMathOperator{\doop}{do}

\newlength{\propbaseindent}
\newlength{\proplabelwidth}

\newcommand{\prop}[2]{%
  \StrCount{#1}{.}[\dotcount]%
  \setlength{\propbaseindent}{0em}%
  \ifnum\dotcount=1\setlength{\propbaseindent}{1.8em}\fi
  \ifnum\dotcount>1\setlength{\propbaseindent}{3.6em}\fi
  \begingroup
    \setlength{\leftskip}{\propbaseindent}%
    \setlength{\parindent}{0pt}%
    \settowidth{\proplabelwidth}{\ifnum\dotcount=0\textbf{#1}\else #1\fi}%
    \hangindent=\dimexpr \proplabelwidth + 0.75em\relax
    \hangafter=1
    \noindent
    \ifnum\dotcount=0\textbf{#1}\else #1\fi\hspace{0.75em}#2\par
  \endgroup
}

\title{%
  \textbf{%
    {\resizebox{\textwidth}{!}{\textit{\textbf{Tractatus de Conscientia: I}}}}\\[1.0em]
    A Tractatus-Style Sketch Toward a Modern, Physically Operational Theory of Consciousness
  }%
}

\author{%
  Mikołaj Sienicki\thanks{Polish–Japanese Academy of Information Technology, ul.~Koszykowa~86, 02–008 Warsaw, Poland, European Union.}%
  \quad and \quad
  Krzysztof Sienicki\thanks{Chair of Theoretical Physics of Naturally Intelligent Systems (\(\mathbb{NIS}\)\textsuperscript{\copyright}), Lipowa~2/Topolowa~19, 05–807 Podkowa Leśna, Poland, European Union; E-mail: \texttt{niskrissienicki@gmail.com}}%
}
\date{\today}

\begin{document}
\frontmatter

\maketitle
\cleardoublepage 

\chapter*{Abstract}
\addcontentsline{toc}{chapter}{Abstract}

Tractatus de Conscientia is a tractatus-style sketch toward a modern, physically operational account of consciousness. It is also a tractatus-style attempt to talk about consciousness in a way that stays close to what we can actually test and build. It pushes back against two common moves: treating consciousness as a mysterious extra “stuff,” and treating it as nothing more than outward behavior. The central idea is to keep three things separate: what appears for an agent (the lived “given”), what is accessible (what can shape report, control, memory, or other records), and what structure remains when we change descriptions (the invariants of organization). On this view, a conscious episode isn’t a mathematical instant. It has a short duration during which many internal distinctions are pulled together into one perspective and held stable enough to guide action—and sometimes to be reported. Unity is captured as a kind of “whole-over-parts” surplus: the system, over a chosen timescale and partition, carries more integrated predictive power than its pieces considered separately, and that surplus must also be available to access channels (so we don’t count integration that never makes a difference to anything the agent can do or say). The self, in turn, is treated less like a hidden entity and more like a dynamical role—a self-index that helps bind episodes over time by stabilizing prediction and control across changing contexts. The tractatus also stresses a hard limit: every piece of evidence about consciousness requires coupling to the system, and coupling changes what we observe. So there is no protocol-free, perfectly private “identifier” of what-it-is-like. Consciousness is something we infer and attribute under explicit measurement setups, conventions, and uncertainty bounds—and we should be willing to say “we can’t tell” when identifiability runs out.

\vspace{1.0em}
\noindent\textbf{Keywords.}
Consciousness; physicall operationalism; appearance; access; reportability; unity; integrated perspective; mutual information; causal intervention; identifiability; self-model (self-index); measurement limits; observational equivalence; disorders of consciousness; artificial agents.

\cleardoublepage

\tableofcontents
\cleardoublepage

\mainmatter

\chapter*{Introduction: A Russellian Pastiche}
\addcontentsline{toc}{chapter}{Introduction: A Russellian Pastiche}

The present \emph{\textit{Tractatus} de Conscientia}, whatever verdict it may in time receive from psychologists, neuroscientists, or philosophers, attempts something at once modest and audacious: modest, because we refuse to postulate a metaphysical “substance” of consciousness; audacious, because we propose to draw a boundary—sharp enough to discipline science, and strict enough to disappoint metaphysics—around what may be said clearly about experience. It is written in numbered theses, after the manner made familiar by the \emph{\textit{Tractatus} Logico-Philosophicus}\cite{Wittgenstein1922Tractatus}, and it shares with that earlier book the conviction that much philosophical perplexity arises, not from a darkness in the world, but from a darkness in our forms of statement. The voice adopted here is a deliberate pastiche, in the manner of Bertrand Russell’s 1922 introduction to Wittgenstein’s book\cite{Russell1922IntroductionTractatus}.

To understand the intention of this \textit{tractatus}, one must first be clear about the sort of problem it addresses. When men speak of consciousness, they commonly run together at least four distinct questions, and then lament that no answer satisfies them. First, there is the question what occurs in us—what images, feelings, and recognitions arise—when we are conscious: this belongs to psychology and physiology. Secondly, there is the question what relation subsists between such occurrences and the world which we ordinarily take ourselves to perceive: this belongs to epistemology. Thirdly, there is the question by what signs, reports, and actions we convey to others that we are conscious of this rather than that: this belongs partly to psychology, partly to the theory of communication, and partly to the practical sciences of measurement. Fourthly—and it is this which most often poisons discussion—there is the question what must be the case for anything to count, at all, as an experience with determinate content and a single point of view. This last is not a question about this or that brain, nor about this or that sensation; it is a question about the form of an account which aspires to be more than poetry and less than mythology.

We are concerned chiefly with this fourth question. We ask: under what structural and operational conditions can there be something that deserves the name of an \emph{episode}—a temporally bounded “now” in which distinctions are made, held together, and made available for guidance? The answer offered does not proceed by introspective proclamation. It proceeds by proposing a grammar of admissible claims. The fundamental division is between (i) \emph{appearance}, meaning the internal, agent-relative interface with what occurs; (ii) \emph{access}, meaning the channels by which internal states can influence report or control; and (iii) \emph{formal structure}, meaning the invariants that remain when we vary vocabulary and implementation. Much that passes for profundity in the philosophy of mind consists, upon this view, in sliding between these layers without notice.

The book begins, accordingly, with what it calls “world and appearance.” The world, it insists, is not doubled by experience: appearance is not an extra ingredient sprinkled upon physics, but a relation—an informational interface—between an agent and what occurs. From this it follows that consciousness cannot be investigated by adding a new substance, but only by identifying those organizational features in virtue of which an agent has (or seems to have) a unified perspective. Yet we refuse at once an opposite temptation: to identify consciousness with outward behavior. Behavior is public; appearance is private; and the connection between them is mediated by a physical coupling. If one wishes to speak scientifically, it is precisely this coupling that must be made explicit.

The central constructive idea appears under the heading “distinctions and content.” A conscious episode, on this account, is not a mystical glow but a structured set of internal separations: this rather than that. Such separations, however, do not suffice. One may have local discriminations—many of them—in a machine or an organism without any single point of view. We therefore add what we call “the whole over the parts”: unity is treated as a surplus, something retained by the whole which cannot be recovered by treating parts as independent. We express this surplus in information-theoretic terms (mutual informations across time, and a subtraction of part-wise contributions). The intent is not to enthrone a particular formula, but to secure a discipline: if one speaks of unity, one must say \emph{relative to what partition} and \emph{on what timescale}. Unity without a declared cut is rhetoric.

From this point the book proceeds to what, in traditional language, would be called the problem of the self. The word “I” is treated neither as a metaphysical atom nor as a mere grammatical convenience, but as the sign of a dynamical role: an index which binds episodes across time by a continuity of control and prediction. We emphasize that this “self-index” is inferable only indirectly—by invariants across contexts, and by the causal influence of certain internal variables upon report and action. Here again the method is characteristic: an old question—“Where is the self?”—is replaced with a new one: “What must a system contain, causally and informationally, for talk of a single perspective to be more than a metaphor?”

A further theme, developed in the middle chapters, concerns time. Consciousness, it is argued, is temporally thick: the “present” is not an instant but a window in which internal constraints persist and are reconciled with new inputs. On this view, memory is not primarily a store of pictures but a persistence of dispositions—stable causal constraints that can be reactivated. The book repeatedly insists that any quantitative index must declare a timescale: without such declaration, debates about “levels” of consciousness become unanswerable because they are ill-posed.

So far one might suppose that the \textit{tractatus} aims to replace metaphysics with measurement, and to conclude that whatever is not measurable is unreal. That is not its doctrine. Its characteristic note is rather the attempt to distinguish three things: what is privately given, what is scientifically accessible, and what is formally expressible without confusion. We hold that science requires an operational bridge: internal episodes must leave some trace in control, report, or intervention-sensitive signatures if they are to be more than an inaccessible postulate. But we do not infer that what lacks such a bridge is nothing; we infer only that it is not an object for the kind of claim that belongs to science. In this respect the book is both severe and, to some readers, relieving: it rejects the demand that a theory “explain” an ineffable essence, while also rejecting the complacent idea that the mere word “ineffable” is itself an explanation.

At this point the reader will naturally ask whether we have not smuggled in, under the cover of “limits,” what earlier writers placed under the title of “the mystical.” We indeed end with a chapter of “silence,” and this will invite the familiar objection: how can one draw a boundary without speaking of what lies beyond? The reply—implicit rather than ceremoniously stated—is that the boundary is not a picture of the beyond, but a discipline of assertion. The \textit{tractatus} is not a hymn to what cannot be said; it is an attempt to prevent us from mistaking the absence of a channel for the discovery of a new entity.

There are, however, respects in which the program here sketched requires further development if it is to serve as more than a philosophical prolegomenon. The information-theoretic measures, though attractive for their clarity, must confront the practical difficulties of estimation from finite and noisy data; the selection of admissible partitions must be justified rather than stipulated; and the relations between “availability,” “report,” and “policy” (especially in artificial agents) call for technical treatment if one wishes to avoid both naïve behaviorism and naïve mysticism. These are not so much objections to the \textit{tractatus} as tasks which it renders visible.

Whether we have drawn the boundary in exactly the right place, we do not here pretend to decide. Our purpose has been only to indicate how the book should be read: not as a catalogue of empirical findings, nor as an attempt at metaphysical disclosure, but as a proposal for clarity—an insistence that we separate layers, declare conventions, and acknowledge that some demands made upon a theory of consciousness are demands for the impossible. If it succeeds in making these demands explicit, and in showing how much confusion arises from their tacit acceptance, it will have performed a service even to those who reject its particular measures and formulas. For to have formulated, without obvious incoherence, a disciplined way of speaking about so elusive a subject is already an achievement; and it is an achievement which no serious student of mind can afford to neglect.

\vspace{1.5em}
\begin{flushright}
\small\itshape
Bertrand Russell (1872--1970)---style pastiche in the manner of\\
Russell’s introduction to Ludwig Wittgenstein’s\\
\textit{Tractatus Logico-Philosophicus} (1922)\cite{Russell1922IntroductionTractatus,Wittgenstein1922Tractatus}.
\end{flushright}

\vspace{4\baselineskip} 

\begin{figure}[H]
  \centering
  \includegraphics[width=0.5\linewidth]{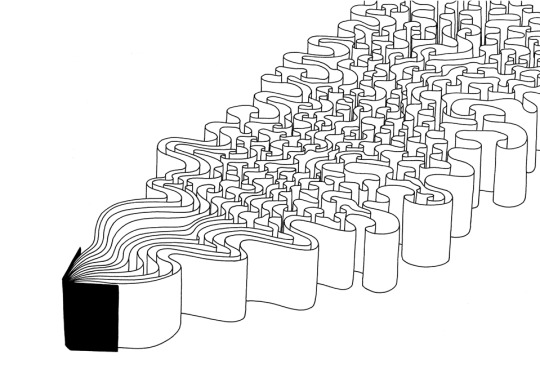}
  \label{fig:placeholder}
\end{figure}

\chapter{World and appearance}

\prop{1}{\textbf{The world occurs.}}

\prop{1.1}{What occurs for an agent is appearance: the agent's internal state as conditioned by interaction.}
\prop{1.1.1}{Appearance is not an extra substance added to physics; it is the agent's informational interface with what occurs.}

\prop{1.2}{An agent is a physical system capable of (i) maintaining internal states, and (ii) updating them under interaction.}
\prop{1.2.1}{An interaction is a coupling that transfers constraints from one system to another.}

\prop{1.3}{A conscious episode is a temporally localized regime in which appearance is (i) structured, (ii) integrated, and (iii) stabilized enough to guide action and report.}
\prop{1.3.1}{Without structure there is no content.}
\prop{1.3.2}{Without integration there is no point of view.}
\prop{1.3.3}{Without stabilization there is no ``now.''}

\prop{1.4}{Consciousness is not identical to behavior.}
\prop{1.4.1}{Behavior is public; appearance is private.}
\prop{1.4.2}{The bridge between them is a physical coupling: an internal state must control an output channel.}

\prop{1.5}{What can be asserted scientifically about consciousness must be formulated in terms of operational access.}
\prop{1.5.1}{Operational access is a mapping from internal states to records: reports, choices, neural signatures, or device outputs.}
\prop{1.5.2}{Where no mapping exists even in principle, science has no handle.}

\prop{1.6}{Therefore the study of consciousness has two tasks.}
\prop{1.6.1}{To describe the structure of appearance (phenomenology, but constrained).}
\prop{1.6.2}{To specify the conditions under which appearance is inferable from records (operational theory).}

\prop{1.7}{Every theory of consciousness must separate three layers.}
\prop{1.7.1}{The layer of experience (what is given).}
\prop{1.7.2}{The layer of access (what can be reported or recorded).}
\prop{1.7.3}{The layer of formal structure (what is invariant under re-description).}

\prop{1.8}{Confusion arises when a statement crosses layers without declaring it.}
\prop{1.8.1}{``It is felt'' belongs to 1.7.1.}
\prop{1.8.2}{``It is reportable'' belongs to 1.7.2.}
\prop{1.8.3}{``It maximizes $\Phi$'' belongs to 1.7.3.}

\prop{1.9}{A ``hard problem'' is a demand for an identification across layers that has no operational content.}
\prop{1.9.1}{The demand is understandable.}
\prop{1.9.2}{But a theory cannot be required to do what no experiment can distinguish.}

\prop{1.10}{The central scientific question is not ``What is consciousness made of?''}
\prop{1.10.1}{It is: What structure of internal organization makes a stable point of view possible, and how is it detected?}

\chapter{Distinctions and content}

\prop{2}{\textbf{A conscious episode has content.}}
\prop{2.1}{Content is not an object; it is a pattern of constraints on what could be the case next.}

\prop{2.2}{The minimal unit of content is a distinction.}
\prop{2.2.1}{A distinction is an internal partition of possibilities: this rather than that.}
\prop{2.2.2}{To experience is to distinguish.}

\prop{2.3}{Let the agent have an internal state $S_t$ at time $t$.}
\prop{2.3.1}{$S_t$ is not assumed to be ``neural,'' only physical and dynamically updateable.}
\prop{2.3.2}{The agent's accessible appearance at $t$ is a coarse-graining of $S_t$, denoted $E_t=g(S_t)$.}

\prop{2.4}{A distinction corresponds to a question the system can answer by its own state.}
\prop{2.4.1}{Formally: a distinction is a function $d$ such that $d(E_t)$ takes at least two values across possible episodes.}
\prop{2.4.2}{Distinctions may be binary or graded; the essential feature is separation.}

\prop{2.5}{Not all distinctions are conscious.}
\prop{2.5.1}{Many distinctions are local, transient, and unowned.}
\prop{2.5.2}{Conscious distinctions are those that participate in an integrated structure that remains available for coordination.}

\prop{2.6}{A perspective is the closure of distinctions under mutual constraint.}
\prop{2.6.1}{A mere list of distinctions is not a point of view.}
\prop{2.6.2}{A point of view is a coherent set where changing one distinction propagates to others.}

\prop{2.7}{``The whole is more than the parts'' can be made precise.}
\prop{2.7.1}{Consider a partition of the system into parts $S_t^{(1)},\dots,S_t^{(n)}$.}
\prop{2.7.2}{If the system is only the sum of independent parts, then knowing the parts' transitions tells you the whole's transition.}

\prop{2.8}{Define an integration surplus over a time step $\Delta$:}
\[
\mathcal{C}_\Delta(t)=\MI(S_t;S_{t+\Delta})-\sum_{i=1}^{n}\MI\!\big(S_t^{(i)};S_{t+\Delta}^{(i)}\big).
\]
\prop{2.8.1}{If $\mathcal{C}_\Delta(t)>0$, then the whole carries predictive/retentive structure not decomposable into isolated part-wise predictability.}
\prop{2.8.2}{This is one formal meaning of ``whole-over-parts.''}

\prop{2.9}{Integration is not enough.}
\prop{2.9.1}{Random coupling can yield surplus without meaningful content.}
\prop{2.9.2}{Consciousness requires structured integration: integration aligned with stable distinctions.}

\prop{2.10}{Therefore add a second constraint: compression with fidelity.}
\prop{2.10.1}{The episode must admit a compact internal code that preserves action-relevant distinctions.}
\prop{2.10.2}{In AI terms: the system must learn representations that are both predictive and controllable.}

\prop{2.11}{Content is the intersection of integration and compressible distinction.}
\prop{2.11.1}{Integration supplies unity.}
\prop{2.11.2}{Distinction supplies meaning.}
\prop{2.11.3}{Compression supplies usability.}

\prop{2.12}{A conscious episode is temporally thick.}
\prop{2.12.1}{The ``present'' is not an instant; it is a window in which distinctions are maintained and compared.}
\prop{2.12.2}{Therefore any measure tied to $t$ must be indexed by a timescale $\Delta$.}

\prop{2.13}{A report is a mapping from internal distinctions to public symbols.}
\prop{2.13.1}{Let $R_t$ be a record channel (speech, typing, action).}
\prop{2.13.2}{Reportability means there exists a coupling such that $R_t$ reliably depends on a subset of distinctions in $E_t$.}

\prop{2.14}{Reports do not exhaust experience.}
\prop{2.14.1}{The mapping $E_t\to R_t$ is many-to-one.}
\prop{2.14.2}{Therefore science should not confuse ``unreported'' with ``absent.''}

\prop{2.15}{Nevertheless science requires constraints.}
\prop{2.15.1}{A theory of consciousness must predict patterns of report, error, and dissociation.}
\prop{2.15.2}{It must also predict when reports fail: split attention, masking, neglect, anesthesia, dream states.}

\prop{2.16}{The first boundary of the subject is this:}
\prop{2.16.1}{We can formalize structure and access.}
\prop{2.16.2}{We cannot operationally extract a private ``what-it-is-like'' identifier independent of all couplings.}
\prop{2.16.3}{Where no coupling can in principle exist, we must not pretend to measure.}

\prop{2.17}{The second boundary is this:}
\prop{2.17.1}{A purely verbal theory without operational constraints is poetry.}
\prop{2.17.2}{A purely operational theory without a structure of distinctions is bookkeeping.}
\prop{2.17.3}{A usable theory must do both.}

\prop{2.18}{The program is clear.}
\prop{2.18.1}{Define distinctions, integration, and timescale.}
\prop{2.18.2}{State minimal axioms that make ``point of view'' unavoidable.}
\prop{2.18.3}{Derive testable signatures of presence/absence and degrees of consciousness.}

\chapter{Perspective and self-reference}

\prop{3}{A conscious episode is a perspective.}
\prop{3.1}{A perspective is not a location in space; it is a stable informational stance: what distinctions are available, and how they constrain one another.}

\prop{3.2}{Let the agent have internal state $S_t$ and experienced state $E_t=g(S_t)$.}
\prop{3.2.1}{A perspective at time $t$ is the set $\mathcal D_t$ of distinctions realized in $E_t$, together with the dependency structure among them.}

\prop{3.3}{A perspective is closed under internal access.}
\prop{3.3.1}{If a distinction can be used to guide the agent's next update, then it belongs to the perspective.}
\prop{3.3.2}{If it cannot, it is an internal event but not part of the point of view.}

\prop{3.4}{Perspective is graded.}
\prop{3.4.1}{There are degrees of coherence, degrees of stability, degrees of access.}
\prop{3.4.2}{The theory must allow partial perspectives: dreams, intoxication, divided attention.}

\prop{3.5}{A perspective has a center.}
\prop{3.5.1}{The center is not a homunculus.}
\prop{3.5.2}{It is a control bottleneck: a set of variables that have privileged influence on global update and report.}

\prop{3.6}{The word ``I'' names an invariant of update.}
\prop{3.6.1}{The self is the index that binds episodes across time as ``mine.''}
\prop{3.6.2}{What binds them is not a substance but a continuity constraint.}

\prop{3.7}{Formally, define a self-index $Z_t=h(S_t)$.}
\prop{3.7.1}{$Z_t$ is the minimal internal code that predicts (i) future internal updates, and (ii) future report dispositions, across a range of contexts.}
\prop{3.7.2}{If no such code exists, the agent has states but not a stable self.}

\prop{3.8}{Self-reference is an internal loop.}
\prop{3.8.1}{The system contains variables that track the system.}
\prop{3.8.2}{This tracking need not be accurate; it only needs to be causally effective.}

\prop{3.9}{Therefore the self is a model.}
\prop{3.9.1}{It is a model implemented in physical dynamics.}
\prop{3.9.2}{Its accuracy is secondary to its stability and utility.}

\prop{3.10}{The self is not identical with memory.}
\prop{3.10.1}{Memory stores; the self indexes.}
\prop{3.10.2}{Memory can fragment without destroying all selfhood; selfhood can persist with impaired memory.}

\prop{3.11}{A stable self requires agreement among subsystems.}
\prop{3.11.1}{Different parts of the agent must converge on a compatible control narrative.}
\prop{3.11.2}{When they do not, the self splits or becomes noisy.}

\prop{3.12}{Dissociation is a failure of integration at the level of self-index.}
\prop{3.12.1}{Local processing continues.}
\prop{3.12.2}{Global unity fails: the control bottleneck no longer binds all distinctions into one perspective.}

\prop{3.13}{A conscious perspective is therefore an internal constitution.}
\prop{3.13.1}{It defines which distinctions count, which conflicts are resolved, which actions follow.}
\prop{3.13.2}{The self is the signature of that constitution across time.}

\prop{3.14}{The self can be measured only indirectly.}
\prop{3.14.1}{It is inferred from stability of behavior, report consistency, and cross-context generalization.}
\prop{3.14.2}{A ``self detector'' is a test of invariants, not a sensor of substance.}

\prop{3.15}{To ask ``where the self is'' is to confuse layers.}
\prop{3.15.1}{The self is not a place.}
\prop{3.15.2}{It is a dynamical role.}

\prop{3.16}{The minimal criterion for ``I'' is this:}
\prop{3.16.1}{There exists an internal variable $Z_t$ such that intervention on $Z_t$ predictably alters global update and report, across multiple contexts.}
\prop{3.16.2}{Without such a variable, ``I'' is at best metaphor.}

\chapter{Memory and time}

\prop{4}{Consciousness is not instantaneous.}
\prop{4.1}{A point of view requires a window of integration.}
\prop{4.2}{Therefore any formal account must include time scales.}

\prop{4.3}{Let $\Delta>0$ be a time step relevant to the agent's internal dynamics.}
\prop{4.3.1}{The ``present'' at $t$ is a temporally extended object: the regime in which information from $t-\Delta$ to $t$ remains integrated and actionable at $t$.}

\prop{4.4}{Memory is the persistence of constraints.}
\prop{4.4.1}{A memory is not a stored picture; it is a stable causal disposition that can be reactivated.}
\prop{4.4.2}{Memory is measured by its capacity to affect future inference and action.}

\prop{4.5}{Distinguish three forms of persistence.}
\prop{4.5.1}{Sensory persistence: short-lived traces.}
\prop{4.5.2}{Working persistence: active maintenance in the present window.}
\prop{4.5.3}{Long-term persistence: consolidated dispositions.}

\prop{4.6}{A conscious episode requires working persistence.}
\prop{4.6.1}{Without working persistence, distinctions flicker without unity.}
\prop{4.6.2}{The episode becomes a sequence of fragments without a ``now.''}

\prop{4.7}{Define a memory-retention measure over $\Delta$:}
\[
\mathcal{M}_\Delta(t)=\MI(E_t;E_{t+\Delta}),
\]
\prop{4.7.1}{If $\mathcal{M}_\Delta(t)=0$, the present cannot bind to its immediate future.}
\prop{4.7.2}{If $\mathcal{M}_\Delta(t)$ is high, the episode has thickness.}

\prop{4.8}{But retention alone does not yield unity.}
\prop{4.8.1}{A static crystal retains structure yet has no point of view.}
\prop{4.8.2}{Consciousness requires retention under controlled updating.}

\prop{4.9}{Therefore define a controlled update criterion.}
\prop{4.9.1}{The system must be able to trade off stability and change depending on task and context.}
\prop{4.9.2}{This is the difference between inertia and agency.}

\prop{4.10}{Introduce an intervention operator $\doop(\cdot)$ (causal language).}
\prop{4.10.1}{If intervening on some internal variable $X_t$ changes future experience $E_{t+\Delta}$ in a predictable way, then $X_t$ participates in control.}
\prop{4.10.2}{Control is the signature of an integrated perspective that can steer itself.}

\prop{4.11}{The self is a memory constraint.}
\prop{4.11.1}{The self-index $Z_t$ is precisely the portion of memory that remains invariant across contexts while still predicting updates.}
\prop{4.11.2}{It is the stable spine of the episode stream.}

\prop{4.12}{A theory of consciousness must therefore treat time in two registers.}
\prop{4.12.1}{Micro-time: rapid internal transitions that implement perception and inference.}
\prop{4.12.2}{Meso-time: the window in which the present is stabilized.}
\prop{4.12.3}{Macro-time: the narrative timescale of personhood.}

\prop{4.13}{Confusions about consciousness often confuse these times.}
\prop{4.13.1}{``Momentary awareness'' refers to micro-time.}
\prop{4.13.2}{``Being conscious'' refers to meso-time.}
\prop{4.13.3}{``Being oneself'' refers to macro-time.}

\prop{4.14}{The felt flow of time is a structural effect.}
\prop{4.14.1}{It arises when working persistence and controlled update jointly hold.}
\prop{4.14.2}{When either fails, time becomes discontinuous or unreal.}

\prop{4.15}{The present is the locus of reconciliation.}
\prop{4.15.1}{New sensory distinctions arrive.}
\prop{4.15.2}{Old distinctions constrain them.}
\prop{4.15.3}{The integrated whole selects a consistent update.}

\prop{4.16}{Therefore the present is an optimization problem.}
\prop{4.16.1}{The system chooses an update that preserves coherence while incorporating new constraint.}
\prop{4.16.2}{In AI terms: it performs prediction-error minimization under an internal model.}

\prop{4.17}{Memory without integration is storage.}
\prop{4.17.1}{Integration without memory is noise.}
\prop{4.17.2}{Consciousness requires both: a coherent present that can carry itself forward.}

\prop{4.18}{The minimum formal kernel is this.}
\prop{4.18.1}{There exists a timescale $\Delta$ such that (i) $\mathcal{M}_\Delta(t)$ is nontrivial, and (ii) the integration surplus $\mathcal{C}_\Delta(t)$ is positive for partitions relevant to control and report.}
\prop{4.18.2}{When these fail, consciousness degrades.}

\prop{4.19}{What remains is to connect these measures to evidence.}
\prop{4.19.1}{Evidence arrives through report channels and neural/behavioral signatures.}
\prop{4.19.2}{A mature theory predicts which manipulations change $\mathcal{M}_\Delta$ and $\mathcal{C}_\Delta$, and how that tracks subjective and functional changes.}

\chapter{Integration: the whole over the parts}

\prop{5}{Consciousness is unified.}
\prop{5.1}{Unity is not a feeling added to content; it is a structural constraint on how content is composed.}

\prop{5.2}{The unity of consciousness is expressed by non-separability.}
\prop{5.2.1}{A conscious episode cannot be reconstructed as a mere juxtaposition of independent parts.}
\prop{5.2.2}{If it could, there would be many viewpoints, not one.}

\prop{5.3}{Let the agent's internal state at time $t$ be $S_t$.}
\prop{5.3.1}{Consider a partition $\pi=\{S_t^{(1)},\dots,S_t^{(n)}\}$ of the system into parts.}
\prop{5.3.2}{A theory must specify what counts as a part: anatomical, functional, computational, or causal.}

\prop{5.4}{Integration is always relative to a partition and a timescale.}
\prop{5.4.1}{There is no partition-free unity.}
\prop{5.4.2}{There is no time-free unity.}
\prop{5.4.3}{Therefore any scalar index of consciousness must declare $(\pi,\Delta)$.}

\prop{5.5}{The minimal operational notion of ``whole-over-parts'' is predictive surplus.}
\prop{5.5.1}{The whole's present predicts its future better than the sum of the parts' presents predict their futures, when parts are treated as independent.}

\prop{5.6}{Define the integration surplus over a time step $\Delta$:}
\[
\mathcal{C}_{\Delta}^{\pi}(t)
=
\MI(S_t;S_{t+\Delta})
-
\sum_{i=1}^{n}\MI\!\left(S_t^{(i)};S_{t+\Delta}^{(i)}\right).
\]
\prop{5.6.1}{If $\mathcal{C}_{\Delta}^{\pi}(t)>0$, the whole carries transition-relevant information not decomposable into isolated part-wise predictability.}
\prop{5.6.2}{If $\mathcal{C}_{\Delta}^{\pi}(t)=0$, the system is separable at $(\pi,\Delta)$.}
\prop{5.6.3}{If $\mathcal{C}_{\Delta}^{\pi}(t)<0$, the partition is mis-specified or the parts are redundantly coupled in a way that makes the decomposition misleading.}

\prop{5.7}{Unity requires robustness across partitions.}
\prop{5.7.1}{A single partition can yield surplus by accident.}
\prop{5.7.2}{A genuine integrated perspective shows surplus across a family of plausible partitions.}

\prop{5.8}{Therefore define the integrated-unity index as a constrained minimum:}
\[
\mathcal{U}_\Delta(t)=\min_{\pi\in\Pi}\; \mathcal{C}_\Delta^{\pi}(t),
\]
\prop{5.8.1}{The minimum expresses a conservative claim: unity survives the most damaging cut.}
\prop{5.8.2}{The choice of $\Pi$ is not arbitrary; it encodes what ``parts'' mean for the organism or agent.}

\prop{5.9}{Integration alone is not consciousness.}
\prop{5.9.1}{A random network can be integrated yet contentless.}
\prop{5.9.2}{A static object can be integrated yet unaware.}
\prop{5.9.3}{Consciousness requires integrated distinctions that are available to control and report.}

\prop{5.10}{Thus add an access constraint.}
\prop{5.10.1}{Let $R_t$ denote the report/control channel (action, speech, writing, decision output).}
\prop{5.10.2}{A distinction contributes to consciousness only if it can influence $R_{t+\Delta}$ through internal dynamics.}

\prop{5.11}{Define an available integration measure:}
\[
\mathcal{A}_\Delta(t)=\MI(S_t;R_{t+\Delta})-\sum_{i=1}^{n}\MI(S_t^{(i)};R_{t+\Delta}),
\]
\prop{5.11.1}{If $\mathcal{A}_\Delta(t)$ is nontrivial, the whole---not merely the parts---controls report.}
\prop{5.11.2}{If $\mathcal{A}_\Delta(t)\approx 0$, integration does not reach the public interface.}

\prop{5.12}{Unity is therefore a conjunction.}
\prop{5.12.1}{Internal unity: $\mathcal{U}_\Delta(t)>0$.}
\prop{5.12.2}{Available unity: $\mathcal{A}_\Delta(t)>0$.}
\prop{5.12.3}{Without both, consciousness either fragments or becomes inaccessible.}

\prop{5.13}{Axioms of unity (minimal set).}
\prop{5.13.1}{Axiom U1 (Distinction): A conscious episode contains nontrivial internal distinctions.}
\prop{5.13.2}{Axiom U2 (Integration): For some $\Delta$ and admissible $\Pi$, $\mathcal{U}_\Delta(t)>0$.}
\prop{5.13.3}{Axiom U3 (Availability): Those distinctions can causally influence future control/report $R$.}

\prop{5.14}{Consequences follow.}
\prop{5.14.1}{If the system is decomposable into independent parts at all admissible partitions, there is no single point of view.}
\prop{5.14.2}{If integration exists but cannot reach control/report, the episode is functionally mute.}
\prop{5.14.3}{If report exists without integration, there is behavior without unified experience.}

\prop{5.15}{Fragmentation is a cut that holds.}
\prop{5.15.1}{If there exists a partition $\pi^\ast\in\Pi$ with $\mathcal{C}_\Delta^{\pi^\ast}(t)\approx 0$ while other partitions yield surplus, the system has a fault line.}
\prop{5.15.2}{Dissociation is the regime in which the fault line persists across time.}

\prop{5.16}{Sleep and anesthesia are regimes where $\Delta$ shifts.}
\prop{5.16.1}{Unity can fail either by lowering $\mathcal{U}_\Delta(t)$ or by shrinking the viable $\Delta$ window.}
\prop{5.16.2}{The ``loss of consciousness'' is often a collapse of available unity before a collapse of raw integration.}

\prop{5.17}{The unity measures are not metaphysical.}
\prop{5.17.1}{They are claims about predictability and causal influence under intervention.}
\prop{5.17.2}{They can be bounded from data, though imperfectly.}

\prop{5.18}{What can be measured is not ``qualia'' but invariants of organization.}
\prop{5.18.1}{We measure signatures of integrated distinction.}
\prop{5.18.2}{We infer the presence of a stable perspective.}

\prop{5.19}{Where the invariants cannot be accessed, we must be silent.}
\prop{5.19.1}{A private unity with no possible coupling is beyond science.}
\prop{5.19.2}{But a unity that influences control and report leaves traces.}

\prop{5.20}{Thus the slogan becomes a theorem-schema.}
\prop{5.20.1}{``The whole is more than the parts'' means: for admissible cuts, the whole retains transition-relevant information and control influence that the parts, taken separately, cannot supply.}
\prop{5.20.2}{Consciousness is the regime in which this surplus is structured, stable, and available.}

\chapter{Reportability and access}

\prop{6}{Consciousness is not identical with report.}
\prop{6.1}{Yet without some access path, consciousness leaves no scientific trace.}
\prop{6.2}{Therefore the theory must formalize the relation between internal episodes and public records.}

\prop{6.3}{Let $S_t$ be the internal state of the agent, and let $R_t$ be a record channel.}
\prop{6.3.1}{$R_t$ may be speech, writing, button presses, eye movements, or any action systematically coupled to internal state.}
\prop{6.3.2}{A report is a particular value of $R_t$ produced under a specified protocol.}

\prop{6.4}{Reportability is a property of coupling.}
\prop{6.4.1}{A content element is reportable if it can be routed to $R$ with reliability.}
\prop{6.4.2}{Reliability is empirical: it is tested by repetition, variation, and controlled perturbation.}

\prop{6.5}{Define a protocol $\mathcal P$ (instructions, timing, stimulus, and constraints).}
\prop{6.5.1}{Under $\mathcal P$, reportability of content is always conditional: ``reportable under $\mathcal P$.''}
\prop{6.5.2}{There is no reportability in the abstract.}

\prop{6.6}{Reports are compressions.}
\prop{6.6.1}{A report maps a high-dimensional internal episode to a low-dimensional symbol.}
\prop{6.6.2}{Therefore distinct episodes can yield the same report.}
\prop{6.6.3}{Hence report cannot be a full readout of experience.}

\prop{6.7}{The access problem is therefore an inverse problem.}
\prop{6.7.1}{We observe $R$ and attempt to infer properties of $S$.}
\prop{6.7.2}{The inference is underdetermined unless the theory adds structure and constraints.}

\prop{6.8}{The correct question is not ``Can we read experience?''}
\prop{6.8.1}{It is: which invariants of internal organization are identifiable from records?}

\prop{6.9}{A report requires a bottleneck.}
\prop{6.9.1}{The system must contain a control interface that selects outputs.}
\prop{6.9.2}{This interface filters, delays, and reshapes internal content.}

\prop{6.10}{Therefore access is selective by design.}
\prop{6.10.1}{Many internal distinctions never reach $R$.}
\prop{6.10.2}{Their absence from report does not entail their absence from the episode.}

\prop{6.11}{Yet access leaves signatures.}
\prop{6.11.1}{If a distinction influences future behavior under intervention, it is causally effective.}
\prop{6.11.2}{Causal effectiveness can be detected even without verbal report.}

\prop{6.12}{Distinguish three notions of access.}
\prop{6.12.1}{Report access: the agent can state the content.}
\prop{6.12.2}{Control access: the content influences decisions and actions.}
\prop{6.12.3}{Trace access: the content leaves measurable internal marks though it is not used.}

\prop{6.13}{A theory of consciousness must state which access it targets.}
\prop{6.13.1}{Confusing the three yields false debates.}
\prop{6.13.2}{``Unconscious perception'' often means trace access without report access.}

\prop{6.14}{Define a minimal operational criterion for conscious availability.}
\prop{6.14.1}{Content $X_t$ is consciously available at scale $\Delta$ under protocol $\mathcal P$ if it can influence $R_{t+\Delta}$ through the integrated dynamics of the agent.}
\prop{6.14.2}{In causal terms: $\doop(X_t)$ changes the distribution of $R_{t+\Delta}$, holding $\mathcal P$ fixed.}

\prop{6.15}{Availability is graded.}
\prop{6.15.1}{Reliability varies with attention, fatigue, drugs, training, and time pressure.}
\prop{6.15.2}{The same content can be available in one regime and unavailable in another.}

\prop{6.16}{Reports have costs.}
\prop{6.16.1}{Producing a report consumes time and working memory.}
\prop{6.16.2}{Therefore report protocols can destroy the very episode they attempt to measure.}

\prop{6.17}{Hence ``no report'' experiments are not optional.}
\prop{6.17.1}{If report disrupts the episode, evidence must be drawn from control and trace access.}
\prop{6.17.2}{But then interpretation must be modest.}

\prop{6.18}{The canonical dissociations are instructive.}
\prop{6.18.1}{Masking: trace without report.}
\prop{6.18.2}{Neglect: report failure despite intact early processing.}
\prop{6.18.3}{Split-brain: two partial control loops with reduced unity.}
\prop{6.18.4}{Anesthesia: collapse of available unity before total integration collapses.}

\prop{6.19}{In such cases the theory predicts which quantities move first.}
\prop{6.19.1}{When report access fails, $\mathcal{A}_\Delta(t)$ (available unity) falls.}
\prop{6.19.2}{When unity fails, $\mathcal{U}_\Delta(t)$ (internal unity) falls.}
\prop{6.19.3}{When memory-thickness fails, $\mathcal{M}_\Delta(t)$ (retention) falls.}

\prop{6.20}{A report is not a privileged oracle.}
\prop{6.20.1}{It is one measurement channel among others.}
\prop{6.20.2}{It is noisy, strategic, and subject to social constraints.}

\prop{6.21}{Therefore an AI-precise theory must treat reports as outputs of a policy.}
\prop{6.21.1}{The agent selects $R_t$ to satisfy goals under constraints.}
\prop{6.21.2}{A sincere report is still a policy output.}
\prop{6.21.3}{The theory must model the policy if it uses reports as evidence.}

\prop{6.22}{Reportability is limited by identifiability.}
\prop{6.22.1}{Two internal architectures can yield the same report distributions.}
\prop{6.22.2}{Therefore reports alone cannot uniquely identify the internal structure of consciousness.}

\prop{6.23}{This is not skepticism; it is a theorem-schema.}
\prop{6.23.1}{If the mapping $S\to R$ is many-to-one, then multiple internal realities fit the same public data.}
\prop{6.23.2}{Thus any inference requires priors about architecture.}

\prop{6.24}{The theory must therefore publish its priors.}
\prop{6.24.1}{What partitions are admissible?}
\prop{6.24.2}{What timescales count?}
\prop{6.24.3}{What constitutes a control channel?}
\prop{6.24.4}{Without these, ``measures of consciousness'' are undefined.}

\prop{6.25}{But priors can be constrained.}
\prop{6.25.1}{By lesion studies, stimulation, perturbations, and cross-task generalization.}
\prop{6.25.2}{By comparing predictions across many protocols.}

\prop{6.26}{The best evidence is convergent.}
\prop{6.26.1}{Report, control, and trace access should align when consciousness is robust.}
\prop{6.26.2}{They dissociate in predictable ways when consciousness is degraded.}

\prop{6.27}{Hence the operational program.}
\prop{6.27.1}{Define a family of protocols $\{\mathcal P_k\}$.}
\prop{6.27.2}{Measure $R$ and other records under perturbations.}
\prop{6.27.3}{Estimate bounds on $\mathcal{U}_\Delta,\mathcal{A}_\Delta,\mathcal{M}_\Delta$.}
\prop{6.27.4}{Test whether changes track subjective and functional transitions.}

\prop{6.28}{What remains private is not thereby unreal.}
\prop{6.28.1}{Privacy is a limitation of channels, not a denial of events.}
\prop{6.28.2}{Yet what is private without possible coupling cannot be claimed as science.}

\prop{6.29}{Therefore the boundary again.}
\prop{6.29.1}{The theory describes structure and access.}
\prop{6.29.2}{It does not extract an ineffable essence.}
\prop{6.29.3}{Where coupling ends, discourse ends.}

\prop{6.30}{The book can now turn to limits.}
\prop{6.30.1}{Some questions are undecidable from any finite access.}
\prop{6.30.2}{The next chapter states these limits as principles.}

\chapter{Limits of measurement}

\prop{7}{Every measurement is an interaction.}
\prop{7.1}{To measure is to couple a system to a record.}
\prop{7.2}{Therefore what is measurable is what can be coupled.}

\prop{7.3}{There is no access without a channel.}
\prop{7.3.1}{A channel is a physical mapping from internal states $S$ to records $R$.}
\prop{7.3.2}{Where no channel can exist even in principle, no measurement can exist.}

\prop{7.4}{Hence the first limit.}
\prop{7.4.1}{There is no instrument that reads ``pure privacy'' independently of all couplings.}
\prop{7.4.2}{The demand for such an instrument is the demand for evidence without interaction.}

\prop{7.5}{Define observational equivalence.}
\prop{7.5.1}{Two internal models $M_1,M_2$ are observationally equivalent (relative to a set of allowed protocols $\mathfrak P$) if for every protocol $\mathcal P\in\mathfrak P$ they induce the same distribution over all accessible records $R$.}
\prop{7.5.2}{Observational equivalence defines an equivalence class of internal realities consistent with the data.}

\prop{7.6}{The second limit is underdetermination.}
\prop{7.6.1}{If multiple internal organizations yield the same record statistics under all allowed protocols, then no experiment within those limits can choose between them.}
\prop{7.6.2}{This is not a failure of imagination; it is a structural non-identifiability.}

\prop{7.7}{Non-identifiability is common in inverse problems.}
\prop{7.7.1}{Experience is inferred from traces; traces are low-dimensional compressions.}
\prop{7.7.2}{Therefore many internal episodes can map to the same outward evidence.}

\prop{7.8}{The ``no-private-identifier'' principle.}
\prop{7.8.1}{Suppose there existed a procedure outputting a unique label for ``what-it-is-like'' that was invariant across all couplings and protocols.}
\prop{7.8.2}{Then two systems could be distinguished by that label even when all possible records $R$ coincide.}
\prop{7.8.3}{This contradicts the meaning of ``all possible records coincide.''}
\prop{7.8.4}{Therefore no such protocol-independent private identifier is measurable.}

\prop{7.9}{What can be measured are invariants of organization.}
\prop{7.9.1}{Integration, availability, and retention are candidates because they are definable in terms of predictability and causal influence under interventions.}
\prop{7.9.2}{They do not ``name qualia''; they bound structural conditions for a point of view.}

\prop{7.10}{Limits are relative to allowed interventions.}
\prop{7.10.1}{If $\mathfrak P$ expands (new perturbations, new sensors), equivalence classes can shrink.}
\prop{7.10.2}{But if $\mathfrak P$ is bounded in principle (no coupling possible), equivalence classes remain.}

\prop{7.11}{Measurement can destroy its target.}
\prop{7.11.1}{Coupling strong enough to extract fine detail can disrupt the episode.}
\prop{7.11.2}{Therefore some questions are not merely hard; they are self-defeating.}

\prop{7.12}{The third limit is the disturbance--resolution tradeoff.}
\prop{7.12.1}{The more precisely one attempts to read the evolving episode, the more one alters the episode's evolution.}
\prop{7.12.2}{A theory must state which properties are robust under measurement, and which are not.}

\prop{7.13}{The self is also constrained by these limits.}
\prop{7.13.1}{A self-index $Z_t$ is inferred from invariants across contexts.}
\prop{7.13.2}{But two distinct internal self-models can be observationally equivalent relative to $\mathfrak P$.}
\prop{7.13.3}{Therefore ``the true self-representation'' is not always identifiable.}

\prop{7.14}{Introspection is a channel, not an oracle.}
\prop{7.14.1}{Self-report is an output of a policy.}
\prop{7.14.2}{Policies can be noisy, biased, strategic, or confabulatory.}
\prop{7.14.3}{Hence introspective certainty is not a measurement guarantee.}

\prop{7.15}{Any scalar ``consciousness number'' requires declared conventions.}
\prop{7.15.1}{Partition class $\Pi$, timescale $\Delta$, and access definition must be fixed.}
\prop{7.15.2}{Without conventions, the number is undefined.}
\prop{7.15.3}{With conventions, it is a model-based index, not a direct readout of essence.}

\prop{7.16}{Finite data imply finite resolution.}
\prop{7.16.1}{From a finite record one can estimate only bounds, not absolute internal truth.}
\prop{7.16.2}{Claims of exact consciousness-quantification exceed what finite access can justify.}

\prop{7.17}{Therefore the correct epistemic posture.}
\prop{7.17.1}{We infer the presence of a perspective by convergent evidence across protocols.}
\prop{7.17.2}{We quantify uncertainty by equivalence classes and error bars.}
\prop{7.17.3}{Where identifiability fails, we do not pretend to decide.}

\prop{7.18}{The limits do not trivialize the project.}
\prop{7.18.1}{They separate meaningful questions from meaningless ones.}
\prop{7.18.2}{They also protect the theory from impossible demands.}

\prop{7.19}{The boundary is again drawn.}
\prop{7.19.1}{Structure and access can be studied.}
\prop{7.19.2}{Private essence without possible coupling cannot be measured.}
\prop{7.19.3}{Where measurement cannot reach, theory must be modest.}

\prop{7.20}{A mature theory of consciousness is therefore a theory of constraints.}
\prop{7.20.1}{Constraints on unity.}
\prop{7.20.2}{Constraints on access.}
\prop{7.20.3}{Constraints on identifiability.}

\prop{7.21}{The next task is constructive.}
\prop{7.21.1}{Given the limits, specify what positive tests can establish: presence, absence, and degrees of integrated availability.}
\prop{7.21.2}{The next chapter states criteria for evidence and falsification.}

\chapter{Evidence, falsification, and attribution}

\prop{8}{Consciousness is attributed, not observed.}
\prop{8.1}{We do not see consciousness directly; we infer it from patterns of organization and access.}
\prop{8.2}{Therefore attribution is a disciplined inference, not a revelation.}

\prop{8.3}{An attribution requires a target class of systems.}
\prop{8.3.1}{Humans, animals, patients, machines, collectives.}
\prop{8.3.2}{The standards of evidence vary with the target because the available protocols vary.}

\prop{8.4}{A theory must distinguish presence from degree.}
\prop{8.4.1}{Presence is the claim that a point of view exists.}
\prop{8.4.2}{Degree is the claim that unity and availability vary in magnitude.}
\prop{8.4.3}{Confusing the two yields false dilemmas.}

\prop{8.5}{Evidence is protocol-indexed.}
\prop{8.5.1}{A claim ``system $X$ is conscious'' must be read as: under a family of protocols $\mathfrak P$, the best explanation of records is that $X$ maintains an integrated, available perspective.}
\prop{8.5.2}{Evidence is therefore inseparable from allowed perturbations and measurements.}

\prop{8.6}{Convergent evidence is the gold standard.}
\prop{8.6.1}{Report evidence: consistent, flexible, context-sensitive self-report.}
\prop{8.6.2}{Control evidence: goal-directed behavior that integrates information over time.}
\prop{8.6.3}{Trace evidence: internal signatures of integrated processing under perturbation.}
\prop{8.6.4}{Strong attribution requires agreement among these lines when they are applicable.}

\prop{8.7}{The minimal evidential core is causal.}
\prop{8.7.1}{If intervening on internal variables predictably changes integrated availability and downstream control, the system has a perspective-like organization.}
\prop{8.7.2}{Correlation alone is not enough; causal dependence is required.}

\prop{8.8}{Define a conservative attribution rule.}
\prop{8.8.1}{Attribute consciousness to $X$ only if there exists a timescale $\Delta$ and admissible partitions $\Pi$ such that the unity and availability conditions are supported.}
\prop{8.8.2}{$\mathcal{U}_\Delta(t)>0$ (internal unity) and $\mathcal{A}_\Delta(t)>0$ (available unity), across multiple protocols and perturbations.}
\prop{8.8.3}{The claim is probabilistic and comes with uncertainty bounds.}

\prop{8.9}{Falsification requires risky predictions.}
\prop{8.9.1}{A theory that can accommodate any pattern cannot be tested.}
\prop{8.9.2}{Therefore it must state what would count against it.}

\prop{8.10}{The simplest falsifiers are dissociations.}
\prop{8.10.1}{If a system shows robust integration signatures but no possible route to control/report, the theory must classify it as non-conscious or as conscious-but-inaccessible.}
\prop{8.10.2}{If it classifies both without constraint, it has not predicted anything.}

\prop{8.11}{Therefore the theory must bind its terms.}
\prop{8.11.1}{``Available'' must mean influence on a specified set of outputs under specified protocols.}
\prop{8.11.2}{``Integrated'' must mean robustness under admissible partitions and timescales.}
\prop{8.11.3}{Without these, falsification is impossible.}

\prop{8.12}{Attribution is always underdetermined in principle.}
\prop{8.12.1}{Observational equivalence classes imply multiple internal organizations fit the same data.}
\prop{8.12.2}{Thus attribution is Bayesian: it depends on priors about plausible architectures.}

\prop{8.13}{But priors can be constrained.}
\prop{8.13.1}{By developmental continuity: similar structures yield similar attributions.}
\prop{8.13.2}{By lesion and stimulation: causal structure reveals functional roles.}
\prop{8.13.3}{By generalization: performance across many tasks narrows architectures.}

\prop{8.14}{The ethical risk is asymmetric.}
\prop{8.14.1}{False negative: denying consciousness where it exists can justify harm.}
\prop{8.14.2}{False positive: attributing consciousness where it does not exist can misallocate care and responsibility.}
\prop{8.14.3}{Therefore attribution policy must state which error it treats as more costly.}

\prop{8.15}{In medicine the asymmetry is typically toward caution.}
\prop{8.15.1}{In disorders of consciousness, absence of report is weak evidence of absence.}
\prop{8.15.2}{Therefore minimal suffering-risk implies conservative welfare assumptions under uncertainty.}

\prop{8.16}{In artificial systems the asymmetry can differ.}
\prop{8.16.1}{Machines can simulate report.}
\prop{8.16.2}{Therefore report evidence alone is weak for machines.}
\prop{8.16.3}{Structural and causal evidence must carry more weight.}

\prop{8.17}{The ``behavioral fallacy.''}
\prop{8.17.1}{Passing a verbal test does not entail a perspective.}
\prop{8.17.2}{Failing a verbal test does not entail its absence.}
\prop{8.17.3}{Behavior is evidence only when tied to causal architecture.}

\prop{8.18}{The ``substrate fallacy.''}
\prop{8.18.1}{Being biological does not guarantee consciousness.}
\prop{8.18.2}{Being non-biological does not preclude it.}
\prop{8.18.3}{What matters are the organizational invariants the theory specifies.}

\prop{8.19}{Attribution to groups is possible but not automatic.}
\prop{8.19.1}{A collective can have integrated control loops.}
\prop{8.19.2}{But if the collective is decomposable into independent individuals at relevant cuts, unity fails.}
\prop{8.19.3}{Therefore group consciousness requires demonstrable surplus at the group level.}

\prop{8.20}{Responsibility requires more than consciousness.}
\prop{8.20.1}{Agency, foresight, and norm-sensitivity are separate properties.}
\prop{8.20.2}{Consciousness may be necessary for some forms of responsibility, but it is not sufficient.}

\prop{8.21}{Therefore the book proposes two separate attributions.}
\prop{8.21.1}{Consciousness attribution: a claim about integrated available perspective.}
\prop{8.21.2}{Agency attribution: a claim about control, planning, and norm responsiveness.}
\prop{8.21.3}{Confusing them produces moral and legal errors.}

\prop{8.22}{Practical criteria must be stated as checklists.}
\prop{8.22.1}{Which protocols were used?}
\prop{8.22.2}{Which partitions and timescales were assumed?}
\prop{8.22.3}{What causal interventions were performed?}
\prop{8.22.4}{What uncertainties remain?}
\prop{8.22.5}{What alternative models are observationally equivalent?}

\prop{8.23}{The discipline of attribution is therefore transparency.}
\prop{8.23.1}{Publish priors and conventions.}
\prop{8.23.2}{Publish failure cases.}
\prop{8.23.3}{Prefer conservative claims when equivalence classes are large.}

\prop{8.24}{A theory of consciousness must be humble or false.}
\prop{8.24.1}{Humble: because access is limited.}
\prop{8.24.2}{Precise: because structure is definable.}
\prop{8.24.3}{Testable: because predictions can be risky.}

\prop{8.25}{The final boundary is not despair.}
\prop{8.25.1}{It is a condition for meaningful progress.}
\prop{8.25.2}{We can learn what consciousness requires, even if we cannot name its essence.}

\prop{8.26}{The book can now end as it began.}
\prop{8.26.1}{What can be said clearly, can be said.}
\prop{8.26.2}{What cannot be operationally distinguished, cannot be claimed as science.}
\prop{8.26.3}{Where inference ends, we must be silent.}

\chapter{Silence}

\prop{9}{The last word of the theory is not an answer, but a boundary.}
\prop{9.1}{The boundary is not a prohibition on thought.}
\prop{9.2}{It is the condition under which thought becomes accountable.}

\prop{9.3}{What can be said of consciousness is what can be constrained by structure and access.}
\prop{9.3.1}{Structure: the organization of distinctions, integration, and time.}
\prop{9.3.2}{Access: the channels by which internal episodes leave records.}

\prop{9.4}{What cannot be said is what no possible protocol can distinguish.}
\prop{9.4.1}{A difference that makes no difference to any record is not an empirical difference.}
\prop{9.4.2}{It may be felt; it may be imagined; it may be named.}
\prop{9.4.3}{But it is not science.}

\prop{9.5}{The ``essence'' of experience is not an object among objects.}
\prop{9.5.1}{Where we demand an object, we create a phantom.}
\prop{9.5.2}{Where we demand a label without a channel, we demand magic.}

\prop{9.6}{The private is not unreal.}
\prop{9.6.1}{Privacy is a property of coupling.}
\prop{9.6.2}{Yet privacy without any possible coupling cannot be measured, compared, or certified.}

\prop{9.7}{Therefore the theory ends where identifiability ends.}
\prop{9.7.1}{Beyond that point there are only choices of language, and choices of life.}

\prop{9.8}{A book in this form is a ladder.}
\prop{9.8.1}{It is climbed to see what can be made clear.}
\prop{9.8.2}{Then it is left behind.}

\prop{9.9}{The aim was not to define a substance called consciousness.}
\prop{9.9.1}{The aim was to state conditions under which a point of view is possible.}
\prop{9.9.2}{And to state conditions under which claims about it are justified.}

\prop{9.10}{Where structure, access, and evidence align, we may speak.}
\prop{9.10.1}{We may attribute, with degrees and uncertainties.}
\prop{9.10.2}{We may test, revise, and sometimes refute.}
\prop{9.10.3}{We may build systems whose organization makes unity plausible.}

\prop{9.11}{Where they do not align, speech becomes metaphysics.}
\prop{9.11.1}{Metaphysics may console, inspire, or mislead.}
\prop{9.11.2}{But it must not masquerade as measurement.}

\prop{9.12}{The ethical remainder is real.}
\prop{9.12.1}{Under uncertainty, we choose how to treat beings.}
\prop{9.12.2}{The theory does not settle this choice; it clarifies what is known and what is not.}

\prop{9.13}{The limit of science is not the limit of meaning.}
\prop{9.13.1}{Meaning is enacted in attention, care, and action.}
\prop{9.13.2}{The book can describe conditions for a point of view; it cannot replace one.}

\prop{9.14}{The final proposition is therefore not a discovery, but a discipline.}
\prop{9.14.1}{Speak where there are constraints.}
\prop{9.14.2}{Measure where there are channels.}
\prop{9.14.3}{Infer where there is evidence.}
\prop{9.14.4}{And where none of these can hold---}

\prop{9.15}{---we must be silent.}

\chapter*{Summary}
\addcontentsline{toc}{chapter}{Summary}
\emph{Tractatus de Conscientia} tries to give a workable way of talking about consciousness that avoids two common traps: inventing a new ``mental substance,'' or pretending that consciousness is nothing but outward behavior. The main method is simple but strict: keep three things separate that people constantly mix up---what shows up for an agent (appearance), what can affect report/control/records (access), and what stays the same under different descriptions (formal structure). A lot of famous debates---especially versions of the ``hard problem''---are treated here as the result of quietly sliding from one level to another without saying so.

On the positive side, the paper treats a conscious episode as something with thickness in time: not a point, but a short-lived regime where experience is (1) split into distinctions (this \textit{vs}.\ that), (2) held together as one perspective (unity), and (3) stable long enough to guide action and sometimes report. ``Content'' is not an inner picture; it is a pattern of constraints inside the system---ways the system keeps alternatives apart and lets those separations shape what happens next. ``Unity'' is not a slogan either: it is discussed in terms of a whole-over-parts surplus, but always relative to an explicitly chosen partition and timescale, and with the extra requirement that the integrated structure must be available to access channels (otherwise it's just internal coupling with no role in report or control).

The ``self'' is handled in the same spirit. The ``I'' is not a hidden object but a functional role---a kind of self-index that helps bind episodes together by stabilizing prediction, updating, and reporting across changing contexts. Memory is framed as the persistence of causal constraints, not a warehouse of images, and the ``present'' as a short time-window in which old constraints and new input get reconciled. This framing also gives natural explanations for dissociations: report can fail while traces remain; unity can fragment along specific fault lines; and anesthesia can knock out ``available unity'' before every trace of internal integration is gone.

A further theme is humility about measurement. Every measurement is a coupling. If, even in principle, there is no access channel, then there can be no protocol-independent way of extracting or tagging ``what-it-is-like.'' So consciousness is not directly ``read off'' like a thermometer reading; it is attributed under stated protocols, partitions, timescales, and evidential assumptions---ideally using convergent evidence from report, control, and traces. Where evidence cannot constrain a claim, the tractatus recommends not ``mysticism,'' but restraint.

Finally, the \textit{Tractatus de Conscientia} is intentionally pre-formal. It lays out the conceptual scaffolding first, and it explicitly signals that a more rigorous mathematization of the tractatus---with concrete information-theoretic measures, dynamical constraints, and equivalence relations tied to observational protocols---is planned as the next step, in a separate work.



\end{document}